\documentclass[pdflatex,sn-aps,iicol]{sn-jnl}

\usepackage{graphicx}%
\usepackage{multirow}%
\usepackage{amsmath,amssymb,amsfonts}%
\usepackage{amsthm}%
\usepackage{mathrsfs}%
\usepackage[title]{appendix}%
\usepackage{xcolor}%
\usepackage{textcomp}%
\usepackage{manyfoot}%
\usepackage{booktabs}%
\usepackage{algorithm}%
\usepackage{algorithmicx}%
\usepackage{algpseudocode}%
\usepackage{listings}%
\usepackage{soul}

\begin{document}

\title[Article Title]{Universal Anyon Tunneling in a Chiral Luttinger Liquid}

\author[1]{\fnm{Ramon} \sur{Guerrero-Suarez}}
\author[1]{\fnm{Adithya} \sur{Suresh}}
\author[1]{\fnm{Tanmay} \sur{Maiti}}
\author[1]{\fnm{Shuang} \sur{Liang}}
\author[1]{\fnm{James} \sur{Nakamura}}
\author[4]{\fnm{Geoffrey} \sur{Gardner}}
\author[5]{\fnm{Claudio} \sur{Chamon}}
\author[1,2,3,4,*]{\fnm{Michael} \sur{Manfra}}
\affil[1]{\orgdiv{Department of Physics and Astronomy}, \orgname{Purdue University}, \city{West Lafayette}, \state{IN}, \postcode{47907}}
\affil[2]{\orgdiv{Elmore Family School of Electrical and Computer Engineering}, \orgname{Purdue University}, \city{West Lafayette}, \state{IN}, \postcode{47907}}
\affil[3]{\orgdiv{School of Materials Engineering}, \orgname{Purdue University}, \city{West Lafayette}, \state{IN}, \postcode{47907}}
\affil[4]{\orgdiv{Microsoft Quantum Lab West Lafayette}, \city{West Lafayette}, \state{IN}, \postcode{47907}}
\affil[5]{\orgdiv{Department of Physics}, \orgname{Boston University}, \city{Boston}, \state{MA}, \postcode{02215}}
\affil[*]{Corresponding author Email: \href{mailto:mmanfra@purdue.edu}{mmanfra@purdue.edu}}%

\abstract{The edge modes of fractional quantum Hall liquids are described by chiral Luttinger liquid theory. Despite many years of experimental investigation fractional quantum Hall edge modes remain enigmatic with significant discrepancies between experimental observations and detailed predictions of chiral Luttinger liquid theory. Here we report measurements of tunneling conductance between counterpropagating edge modes at $\nu=1/3$ across a quantum point contact fabricated on an AlGaAs/GaAs heterostructure designed to promote a sharp confinement potential. We present evidence for tunneling of {\it anyons} through a $\nu=1/3$ incompressible liquid that exhibits universal scaling behavior with respect to temperature, source-drain bias, and barrier transmission, as originally proposed by Wen \cite{Wen1990,Wen1991}. For transmission $t\geq0.800$, we measured the tunneling exponent $\bar{g} = 0.333 \pm 0.005$ averaged over 29 independent data sets, consistent with the scaling dimension $\Delta = g/2 = 1/6$ for a Laughlin quasiparticle at the edge. When combined with measurements of the fractional charge $e^*=e/3$ and the recently observed anyonic statistical angle $\theta_a=\frac{2\pi}{3}$, the measured tunneling exponent fully characterizes the topological order of the primary Laughlin state at $\nu=1/3$.}

\maketitle
 
\section*{Introduction}

Three quantities - fractional charge $e^*$~\cite{Laughlin1983}, anyonic statistical angle $\theta_a$~\cite{Halperin1983,Arovas1984}, and the tunneling exponent $g$~\cite{Wen1990,Wen1991}, provide a universal parameterization of the topological order \cite{Wen1995} of Abelian fractional quantum Hall effect (FQHE) liquids in the Laughlin sequence. The tunneling exponent $g$ is connected to the scaling dimension, $\Delta=g/2$, of the quasiparticles that propagate at the boundary of an incompressible FQHE liquid \cite{Wen1990,Wen1991,Wen1995,Kane1992PRL,Kane1992,Kane1994,Fendley1994,Fendley1995PRL,Fendley1995PRB, ChamonWen1992, ChamonFreedWen1995}. For Laughlin states, $g=\nu^{-1}$ for electrons and $g=\nu$ for fractional quasiparticles, where $\nu$ is the filling factor of the incompressible state. Although theoretical predictions for these fundamental quantities have been available for more than three decades, experimental validation has proceeded more slowly. Fractionalization of charge was first observed in 1997 following advances in quantum shot noise measurements~\cite{dePicciotto:1997qc,Glattli1997}. Anyonic braiding statistics were demonstrated in 2020 using electronic Fabry-Pérot interferometers~\cite{Nakamura2020, Nakamura2022, Nakamura2023}, and with noise correlation measurements in novel anyon collider experiments~\cite{Bartolomei2020, Ruelle2023PRX, Lee2023Nature}. Despite these advances, experimental observation of tunneling of anyons and the expected scaling behavior of the chiral Luttinger liquid predicted to exist at the boundary of FQHE liquids have proven to be more elusive. It is theoretically well-established that the properties of the edge modes are a sensitive probe of the bulk topological order, thus reconciling experimental observations with theory remains of paramount importance for full understanding and categorization of FQHE liquid phases. 

Counterpropagating edge modes of a fractional quantum Hall liquid may be brought into close proximity to induce tunneling with a quantum point contact (QPC). Early experiments with QPCs in the fractional quantum Hall regime explored behavior with the constriction set to very low transmission \cite{Milliken1996SSC}, the so-called strong backscattering regime. In the strong backscattering limit, the tunneling between the edge modes is mediated by {\it electrons}. In experiments utilizing elegant cleaved edge overgrowth heterostructures, Chang and Grayson ~\cite{Chang1996PRL,Grayson1998PRL,Chang2001PRL} explored tunneling of electrons to the edge of a fractional quantum Hall state from a 3D Fermi gas separated from the edge of a fractional quantum Hall liquid by an AlGaAs tunnel barrier. Although these and other early resonant tunneling experiments \cite{GoldmanSu1995, Maasilta1997PRB} gave indications of the impact of strong correlation effects and deviations from the known properties of Fermi liquids, they also revealed important discrepancies with the predictions of chiral Luttinger liquid theory, most notably in the filling factor dependence of the tunneling exponent. 

In the weak backscattering limit, when a QPC sufficiently narrows the distance between counterpropagating edge modes that bound an incompressible quantum liquid, weak tunneling between counterpropagating edge modes is expected to be mediated by fractionalized (anyonic) quasiparticles. Previous experiments in the weak tunneling regime at $\nu = 1/3$ have shown qualitative and quantitative disagreement with the behavior expected from chiral Luttinger liquid theory~\cite{Roddaro2003,Roddaro2004PRL,Roddaro2004SolidState,Baer2014,Ensslin2018}. Perhaps the most indicative signature of the qualitative inconsistency between experiment and theory was the persistent observation of a {\it minimum} in tunneling conductance at low bias and lowest temperatures, while theory predicts a {\it maximum} in zero bias tunneling conductance for a chiral Luttinger liquid at $\nu=1/3$. Qualitative and quantitative differences between experiments and theory remain a fundamental challenge to our understanding of the strongly interacting one-dimensional liquid that forms the boundary of FQHE states. It is important to establish the bulk-boundary correspondence in Abelian Laughlin states if edge-mode dynamics is to be used to probe more exotic non-Abelian topological order \cite{Miller2007Nature,Radu2008}.

Recent experimental developments give reason for optimism. Veillon {\it et al.} \cite{veillon2024observation} reported on the determination of the scaling dimension of Abelian FQHE liquids by examination of the crossover from thermal noise to shot noise following a recent theoretical analysis \cite{Snizhko2015,Schiller2022}. Germane to the work discussed here, an impressive scaling behavior of conductance was recently demonstrated in a graphene device operating in the quantum Hall regime~\cite{Cohen2023}. This innovative experiment measured the tunneling of {\it electrons} into a $\nu = 1/3$ edge mode from a counterpropagating $\nu=1$ edge mode. Gates were used to tune the local electron density so that the regions of $\nu=1$ and $\nu=1/3$ were brought in close proximity to allow electron tunneling. 

In this work, we present evidence of direct tunneling of {\it anyons} across a droplet of $\nu=1/3$ liquid and demonstrate universal scaling behavior consistent with the theoretical description of a chiral Luttinger liquid. We focus on the weak backscattering regime, where quantitative comparisons with theory are readily made, and we extract $\bar{g}=0.333~\pm 0.005$, for the tunneling exponent for the $\nu=1/3$ state for transmission $t\geq0.800$ at $T_e=34$~mK. Furthermore, we demonstrate scaling behavior of the tunneling conductance with respect to temperature, source-drain bias, and barrier transmission as originally outlined by Wen and collaborators~\cite{Wen1990, Wen1991, ChamonWen1992, ChamonWen1994}. 

Our starting point for data analysis is the universal functional form of the tunneling current originally derived by Wen~\cite{Wen1991}. This formulation was derived using perturbation theory assuming a small tunnel coupling between counterpropagating edge modes, and yields the tunneling differential conductance
\begin{align}\label{eq.1}
G_t = \frac{e^2}{h}\left(\frac{2\pi T}{T_0}\right)^{2g-2}f_g\left(e^* \frac{V_{SD}}{k_B T}\right)
\end{align}
where $T$ is the electron temperature, $T_0$ is an effective temperature scale proportional to the tunnel coupling between the edge modes, and $V_{SD}$ is the applied source-drain bias. $e^*$ is the fractional charge of the quasiparticle that undergoes tunneling and $g$ is the tunneling exponent. The scaling function $f_g(x)$ is:
\begin{align}\label{eq.1a}
f_g(x) = &B\left(g+i\frac{x}{2\pi},\, g-i\frac{x}{2\pi}\right) \;\cosh\left({x/2}\right)\nonumber\\
   &\times\left\{\pi - 2\tanh\left({x/2}\right)\,\mathrm{Im}\left[\psi\left(g+i\frac{x}{2\pi}\right)\right]\right\}
\end{align}
where $B(a, b)$ is the beta function and $\psi(x)$ is the digamma function, see Supplementary Information for more details. This universal scaling form, along with the constraints from the chiral Luttinger liquid theory that $e^*=\nu\,e$ and $g=\nu$ for anyon tunneling, result in sharp predictions for measurements of the tunneling conductance.
\begin{figure*}
\centering
\includegraphics[width = \linewidth]{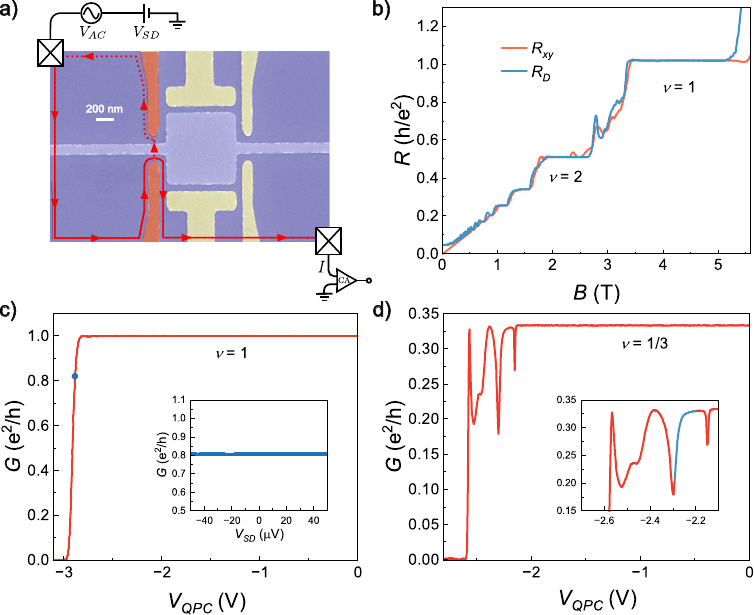}
\caption{\raggedright \textbf{(a)} False color scanning electron microscopy image of a full interferometer with a schematic of the measurement circuit used in this experiment. The QPC used for tunneling measurements is highlighted in orange. All other gates are grounded. Red lines represent edge mode circulation. \textbf{(b)} Simultaneous measurement of bulk $R_{xy}$ and $R_D$ across the QPC as a function of magnetic field at $T_{mc}=10$~mK. The QPC is biased just past depletion to define the current path. The filling factor is the same in the QPC as in the bulk of the Hall bar. \textbf{(c)} QPC conductance at the center of the $\nu = 1$ ($B = 4.20$~T) plateau. The conductance is quantized to $e^2/h$ over most of the voltage range and exhibits a sharp pinch-off. This data indicates a sharp confining potential in our QPC built upon the screening well heterostructure design. The inset displays differential conductance at $\nu = 1$ taken with $t\approx0.8$ as indicated by the blue point on the conductance plot. This constant differential conductance is expected for the $\nu = 1$ edge described by Fermi liquid theory, in sharp contrast to the behavior observed at $\nu = 1/3$. \textbf{(d)} QPC conductance at $\nu = 1/3$. The conductance is quantized to $e^2/3h$ over most of the voltage range and displays a sharp pinch-off. Unlike the behavior observed at $\nu = 1$, a few sharp resonances are observed at $\nu = 1/3$ prior to full depletion. The inset highlights the region where tunneling was investigated.}
    \label{Fig1}
\end{figure*}
\section*{Device Configuration and Measurement Results}
\begin{figure*}
    \centering
    \includegraphics[width = \linewidth]{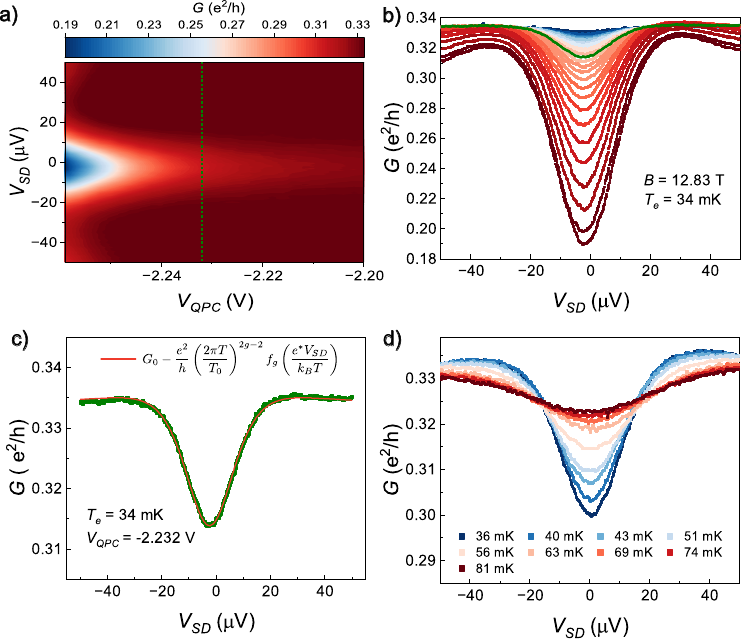}
   \caption{\raggedright \textbf{(a)} 2D color map of differential conductance plotted as a function of QPC voltage $V_{QPC}$ and DC source-drain bias $V_{SD}$. \textbf{(b)} Line cuts of differential conductance measured at several QPC transmissions at fixed electron temperature $T_{e}$=34 mK at the center of the $\nu = 1/3$ plateau. The transmission through the QPC varies from $t \approx 0.99$ to $t \approx 0.65$. The linecuts are extracted from the data of Fig.~2a. \textbf{(c)}  Differential conductance at $t = 0.94$. The red line shows the fit to the data using the expression for the tunneling conductance in the weak backscattering limit derived by Wen and collaborators~\cite{Wen1990, Wen1991, ChamonWen1992, ChamonWen1994}. \textbf{(d)} Temperature dependence of the differential conductance for QPC transmission $t = 0.9$. Temperatures listed in the legend are mixing chamber temperatures at which each data set was collected.}
    
    \label{Fig2}
\end{figure*}
The device studied in this experiment utilizes a quantum point contact which is a functional component of a Fabry-P{\'e}rot interferometer~\cite{Chamon1997}. This interferometer was previously used to demonstrate anyonic braiding statistics in the fractional quantum Hall effect at filling factor $\nu=2/5$ \cite{Nakamura2023}. It has two QPCs and a pair of plunger gates that define the interference path when in operation. The arms of the QPC are separated by $300\; \text{nm}$ and the plunger gates are $1\; \mu \text{m}$ apart. A false color scanning electron microscopy image of a nearly identical device with a schematic of the measurement circuit is shown in Fig.~1a. Only the highlighted QPC is energized to study tunneling phenomena, while all other gates are held at ground potential, including the large gate positioned above the interior of the interferometer. Importantly, this device was fabricated on an AlGaAs/GaAs heterostructure that we identify as the screening well design \cite{Nakamura2019, Nakamura2020, Nakamura2022, Nakamura2023, Shuang2025}. Two additional quantum wells placed above and below the principal quantum well are populated with high-density two-dimensional electron gases. These screening wells, situated 25~nm above and below the principal quantum well, create a sharp confinement potential at the location of the QPC in the central well; sharp confinement produces high edge mode velocity and is critical to the prevention of edge reconstruction~\cite{ChamonWen1994} that may otherwise confound interpretation of tunneling data. This characteristic may account for the differences between the results presented here and previous experimental investigations of tunneling using QPCs in the fractional quantum Hall regime in AlGaAs/GaAs heterostructures~\cite{Umansky2019, Ensslin2018, Roddaro2004PRL, Roddaro2004SolidState, Roddaro2003, GoldmanSu1995, schiller2024scaling, veillon2024observation}. To avoid parallel conduction through the screening wells, we utilize depletion gates above and below the arms connecting the ohmic contacts to the interferometer~\cite{EisensteinAPL1990}. These gates are negatively biased to locally deplete the top and bottom screening wells in a small region around the contacts, but leave the main quantum well fully populated and connected to the ohmic leads (see Supplementary Information and Fig.~S1b for more details). The electron density of the main quantum well is $n=1.0\times 10^{11}$cm$^{-2}$ and its mobility is $\mu = 9 \times 10^6$cm$^2$/Vs in the fully fabricated device geometry. All conductance measurements are made with standard lock-in techniques with an AC excitation bias of 2~$\mu$V and an excitation frequency of $11~\text{Hz}$. We measure the AC current transmitted through the QPC while varying device parameters including DC source-drain bias $V_{SD}$, QPC transmission $t$, electron temperature $T_e$, and applied magnetic field $B$. The differential conductance $G=\frac{\partial I}{\partial V}\vline_{V_{SD}}$ is the conductance measured at specific values of DC source-drain bias $V_{SD}$. The tunneling conductance $G_t=\frac{\partial I_t}{\partial V}\vline_{V_{SD}}$ measures the current scattered to the counterpropagating edge across the QPC. In our measurement configuration, the differential conductance $G$ is directly measured and the tunneling conductance is given by: $G_t=G_0-G$ with $G_0=\frac{e^2}{3h}$ at $\nu=1/3$ and $G_0=\frac{e^2}{h}$ at $\nu=1$. We study tunneling between counterpropagating edges at filling fractions $\nu=1$ and $\nu=1/3$ in the weak tunneling regime, defined as $G_t \ll \sigma_{xy}$ where $\sigma_{xy}$ is the Hall conductance. In the limit of sharp confining potential, only a single chiral edge mode is observed at $\nu=1$ and $\nu=1/3$. 

Fig.~1b displays magnetoransport through the device. We simultaneously measure the bulk Hall resistance $R_{xy}$ measured on the mesa but away from the QPC, and the diagonal resistance $R_D$ measured across the QPC. This measurement is performed with $V_{QPC} = -0.4\,$V, just past the depletion point of the principal quantum well. As seen in Fig.~1b, $R_{xy}$ and $R_{D}$ coincide at all integer quantum Hall states, indicating that the filling fraction in the QPC is the same as the bulk filling fraction and supports incompressible states within the QPC. The centers of the $\nu=1$ and $\nu=1/3$ plateaus are located at $B= 4.20\,$ T and $B= 12.83\,$ T, respectively, see Supplementary Information for additional details.
\par
In the weak backscattering regime, tunneling between counterpropagating edge modes is mediated by electrons at $\nu=1$ and by fractionalized anyons across a gapped fractional quantum Hall liquid at $\nu=1/3$. Fig.~1c displays the conductance with zero DC source-drain bias as a function of the QPC voltage at the center of the $\nu = 1$ plateau. We then set the QPC voltage to partially reflect the single $\nu=1$ edge mode with transmission $t \approx 0.8$ and measure the differential conductance as a function of $V_{SD}$. The edge mode at $\nu=1$ is expected to behave as a normal Fermi liquid of electrons. Ohmic behavior is expected with constant differential conductance, consistent with the data displayed in the inset of Fig.~1c. With confirmation of ohmic behavior at $\nu=1$, we then raise the magnetic field to B=12.83\,T, the center of the $\nu = 1/3$ plateau. As evident in Fig.~1d, the conductance through the QPC remains quantized over most of the QPC voltage range in a manner similar to that observed at $\nu = 1$, but also exhibits a few sharp resonances before full depletion. These resonances are a fingerprint of the specific mesoscopic disorder configuration at $\nu=1/3$ in this QPC. The inset to Fig.~1d highlights the region where we measure differential conductance in the weak tunneling regime. The sharp pinch-off observed at both $\nu = 1$ and $\nu = 1/3$ indicates that the QPC produces a sharp confinement potential consistent with the screening well heterostructure design and our previous interferometry results \cite{Nakamura2019, Nakamura2020, Nakamura2022, Nakamura2023}. If the edge confining potential is not sharp, the edge mode velocity is reduced and edge mode reconstruction is possible~\cite{Umansky2019, Rosenow2002, Papa2004}. Edge mode reconstruction may be the root cause of the long-standing discrepancy between predictions of chiral Luttinger liquid theory and previous experimental studies of quasiparticle tunneling~\cite{Umansky2019, schiller2024scaling, Ensslin2018, Roddaro2003, Roddaro2004PRL}. 

The edge at $\nu = 1/3$ is described as a single chiral charge mode~\cite{Wen1990, Wen1991, Kane1992}. The tunneling conductance in the weak backscattering limit has been calculated in perturbation theory and is given by~\eqref{eq.1}. For the $\nu=1/3$ Laughlin state, it is expected that the scaling exponent $g$ in equation~\eqref{eq.1} will be quantized to the universal value $g=1/3$. In the scaling function~\eqref{eq.1}, $T_0$ is an effective temperature that represents the strength of anyon tunneling and is proportional to the QPC transmission determined by the voltage applied to the gates, $V_{QPC}$. Fig.~2a displays a 2D color map of the differential conductance $G$ while the QPC transmission is varied by tuning $V_{QPC}$ at $B=12.83$~T and $T_e=34$~mK. We note that $T_e$ refers to the electron temperature measured concurrently by Coulomb blockade thermometry using the full interferometer as a quantum dot. This approach allows us to accurately measure the electronic temperature in the operating device. The QPC transmission is varied from $t \approx 0.99$ to $t \approx 0.65$. Importantly, the dependence of the conductance on $V_{SD}$ at $\nu=1/3$ displays the expected functional dependence predicted by chiral Luttinger liquid theory in the weak tunneling limit. This behavior is evident in the individual line cuts displayed in Fig. 2b. At high bias the differential conductance saturates to the quantized value of $e^2/3h$. The zero bias ($V_{SD}=0$) conductance also displays a pronounced minimum. To the best of our knowledge, this qualitative behavior of a minimum in differential conductance at zero DC bias and saturation to $G=\frac{e^2}{3\hbar}$ has not been observed previously in experiments in which tunneling between fractional quantum Hall edge modes was studied in the weak tunneling limit at $\nu=1/3$~\cite{Roddaro2003,Roddaro2004PRL,veillon2024observation}. 
\par
Our data not only reproduces the qualitative behavior predicted by chiral Luttinger liquid theory, but significantly, facilitates extraction of the scaling exponent with high precision. Fig.~2c displays the differential conductance with $t=0.94$ at $T_e=34$~mK along with a least squares fit to the expression for the differential conductance, $G=G_0-G_t$. In our fit $G_0$, $T_0$ and $g$ are left as free parameters. $e^*$ is set to $e^* = \frac{e}{3}$ and $T_{e}=34~$mK. The effective quasiparticle charge of $\frac{e}{3}$ is independently measured by operating this device as a Fabry-P\'erot interferometer at $\nu=1/3$~\cite{Nakamura2023, Nakamura2019, Nakamura2020, Nakamura2022}. The value extracted for the scaling exponent for the specific data set shown in Fig.~2c is $g=0.334$ with $T_0=1.34$~mK and $G_0=0.334$, see Supplementary Information for a more detailed description of the fitting procedure and residual uncertainties. The extraction of $g$ and $G_0$ with values close to the expectations for a chiral Luttinger liquid at $\nu=1/3$ is robust against changes of QPC transmission $t$, controlled by $V_{QPC}$. Although only single data set and corresponding fit are displayed in Fig.~2c, we fit all 35 independent data sets of Fig.~2b. The values of $g$, $G_0$ and $T_0$ as a function of the transmission $t$ are shown in Supplementary Information Fig.~S2. For transmissions $0.800\leq t\leq 0.995$, the average of $g$ and $G_0$ determined from 29 independent data sets are $\bar{g}=0.333\pm0.005$ and $\bar{G_0}=0.333\pm0.001\text{ }e^2/h$. Here the standard deviation is reported as the uncertainty.

In Fig.~3a we plot the experimentally determined zero bias tunneling conductance $G_t=\frac{\partial I_t}{\partial V}\vert_{V_{SD}=0}$ as a function of the dimensionless parameter $T/T_0$. For each data set displayed in Fig.~2b, we extract a value of $T_0$. These values along with $T_e=34$~mK determined by {\it in operando} Coulomb blockade thermometry set the ratio $T/T_0$. The inset to Fig.~3a displays the same data plotted on a log-log scale. The linear dependence of $G_t$ over nearly two orders of magnitude clearly demonstrates the power law dependence expected for the tunneling of anyons in a chiral Luttinger liquid. Also included in Fig.~3a are plots of the theoretical dependence of $G_t$ on $T/T_0$ when one {\it assumes} different values for $g$ with $g=0.3,1/3,0.4$ displayed for comparison.

We extend the range over which the dependence of $G_t$ on $T/T_0$ is probed by repeating our measurement and fitting procedure at two additional electron temperatures, $T_e = 47\,\text{mK}$ and $T_e = 69\,\text{mK}$. These results are shown in Fig.~3b. We observe that the data agree well with the expected behavior for precisely $g = 1/3$ over a broad range in $T/T_0$ for these data sets as well.

The quantization of $g=1/3$ is expected to be maintained for deviations of the magnetic field away from the center of the $\nu=1/3$ plateau if the state remains incompressible. We repeated our measurements at different magnetic field values along the $\nu = 1/3$ plateau, $\Delta B=\pm 90\,\text{mT}$ away from the center of the plateau. As seen in Fig.~3c, the data continue to show good agreement with the theoretical prediction $g = 1/3$.

It can be seen in equation~\eqref{eq.1} that the tunneling conductance is a universal function of the dimensionless parameter $e^*V_{SD}/k_B T$.
\begin{figure}
    \centering
    \includegraphics[width = 80mm]{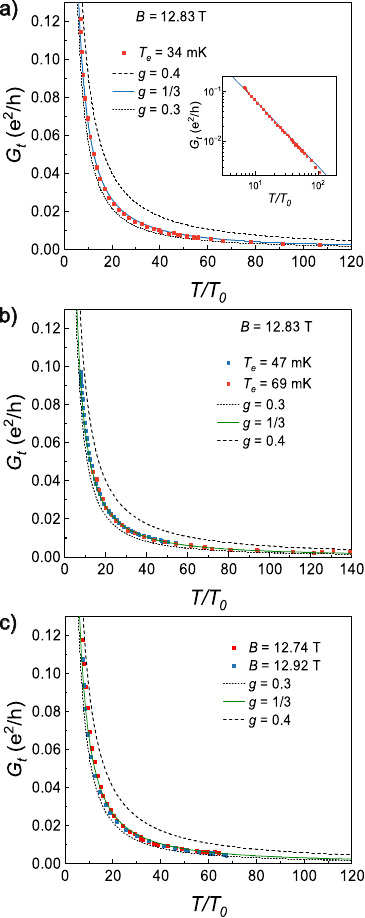}
    \caption{\textbf{(a)} Zero bias tunneling conductance vs $T/T_0$ at $T_e = 34\,\text{mK}$ and $B=12.83$~T, at the center of the $\nu = 1/3$ plateau. Each point is extracted from a fit of each trace in Fig.~2b as in Fig~2c. The lines correspond to the theoretical expectation at $g = 1/3$, $g = 0.3$ and $g = 0.4$. The inset shows the data plotted on a log-log scale. \textbf{(b)} Zero bias tunneling conductance vs $T/T_0$ at $T_e = 47\,\text{mK}$ and $T_e = 69\,\text{mK}$, at the center of $\nu = 1/3$. \textbf{(c)} Zero bias tunneling conductance vs $T/T_0$ at $B = 12.92\, \text{T}$ and $B = 12.74\, \text{T}$, which are $\Delta B\pm 90\,~\text{mT}$ away from the center of the $\nu = 1/3$ plateau.
}
    \label{Fig3}
\end{figure}
To demonstrate universal scaling in $e^*V/k_B T$, we fixed the QPC voltage and measured the differential conductance at several different temperatures. This data is shown in Fig.~2d for QPC transmission of $t = 0.9$ and dilution refrigerator temperature ranging from $T_{MC} = 36\,\text{mK}$ to $T_{MC} = 81\,\text{mK}$. For these data sets, we report the mixing chamber temperature, as we do not perform {\it in operando} CB thermometry at each temperature setting reported here. For mixing chamber chamber temperatures $T_{MC}\geq40$mK, calibration experiments indicate $T_e=T_{MC}$. To test the scaling behavior, we define $\tilde{G}$:
\begin{align}\label{eq. 2}
    \tilde{G} = \frac{h}{e^2}\left(\frac{T_0}{2\pi T}\right)^{2g-2}\left(G_0-G\right)
\end{align}
where the tunneling conductance, $G_0 - G$, is scaled by a universal coefficient. After rendering the conductance data in Fig.~2d in this manner and plotting $\tilde G$ as a function of $e^*V_{SD}/k_B T$, we observe the scaling collapse displayed in Fig.~4a and b. In Fig.~4a, $\tilde{G}$ is plotted on a linear scale versus $e^*V_{SD}/k_BT$. The reduced conductance at different temperatures collapses into a single curve given by the universal function:
\begin{align}
    \tilde{G} = f_g\left(e^* \frac{V_{SD}}{k_B T}\right)
\end{align}
The scaling collapse becomes more evident when $\tilde{G}$ versus $e^*V_{SD}/k_BT$ is plotted on a log-log scale as displayed in Fig.~4b. 
\begin{figure}
    \centering
    \includegraphics[width = 80mm]{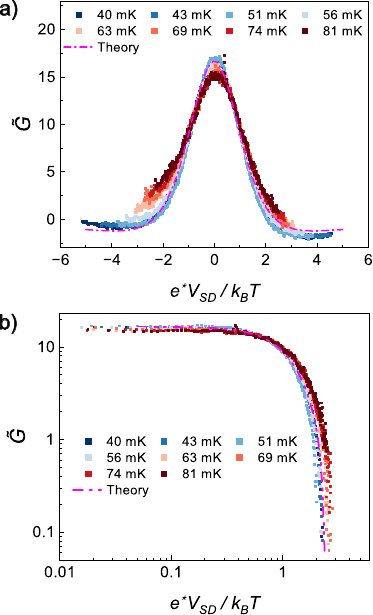}
    \caption{\raggedright \textbf{(a)} Scaling behavior of reduced tunneling conductance $\tilde{G}$ vs. $e^*V_{SD}/k_B\,T$ using data of Fig.~2d. Data collected at different temperatures collapse onto a single curve. Dashed magenta line is the theoretical expectation for $g=1/3$.  \textbf{(b)} Data in (a) plotted on log-log scale. Dashed magenta line is the theoretical expectation for $g=1/3$.
 }
    \label{Fig4}
\end{figure}

\par
\section*{Discussion}
The data plotted in Figs.~4a and 4b demonstrate that the tunneling conductance obeys robust scaling in the dimensionless variable $e^*V_{SD}/k_B T$. In particular, the scaling behavior follows the predictions of chiral Luttinger liquid theory for the $\nu=1/3$ state in the weak backscattering limit. The theoretical expectation is indicated by the dashed magenta lines in Figs.~4a and 4b that overlap with the experimental data. One may ask why this behavior has not previously been observed in AlGaAs/GaAs heterostructures used in the study of the fractional quantum Hall effect. The most significant difference between the experiments reported here and previous generations of experiments employing AlGaAs/GaAs heterostructures is the introduction of the screening well heterostructure into our device architecture. Ancillary screening wells have been effectively used to control Coulomb charging effects in the operation of Fabry-P{\'e}rot interferometers, and have also been shown to provide a sharper confining potential at the edge of the FQHE liquid \cite{Nakamura2019}. We speculate that the close agreement with the theoretical predictions seen here is due to the absence of edge reconstruction that is typically observed with standard heterostructures. In addition, the filling factor in the QPC is well defined and matches that in the bulk, providing a droplet of incompressible liquid supporting anyon excitations. Under these conditions, the correspondence with predictions for a single chiral Luttinger liquid edge mode is manifest.
\par
\section*{Conclusions}
Using a QPC built on an AlGaAs/GaAs screening well heterostructure, we measured tunneling conductance between counterpropagating edge modes of the $\nu=1/3$ FQHE state in the weak backscattering limit. Our data facilitate the extraction of the tunneling exponent $\bar{g}=0.333\pm0.005$, consistent with expectations for anyon tunneling in a chiral Luttinger liquid at $\nu=1/3$. In addition, we observed scaling behavior in the dimensionless variable $e^*V_{SD}/k_BT$. These findings indicate that the classification of the topological order responsible for the bulk quantization of FQHE liquids can be fully characterized by a sequence of measurements with a Fabry-P{\'e}rot interferometer. 

\section*{Acknowledgments}
This research is sponsored by the U.S. Department of Energy, Office of Science, Office of Basic Energy Sciences, under the award numbers DE-SC0020138 and DE-FG02-06ER46316. The content of the information presented here does not necessarily reflect the position or the policy of the US government, and no official endorsement should be inferred.

\bibliography{sn-bibliography}

\clearpage
\backmatter

\begin{appendices}
\renewcommand{\thefigure}{S\arabic{figure}}
\renewcommand{\theequation}{S\arabic{equation}}
\setcounter{figure}{0}
\setcounter{equation}{0}
\section*{Supplementary Information}
\subsection*{AlGaAs/GaAs Screening Well Heterostructure and Device Operation}
\begin{figure}[htp!]
    \centering
    \includegraphics[width = \linewidth]{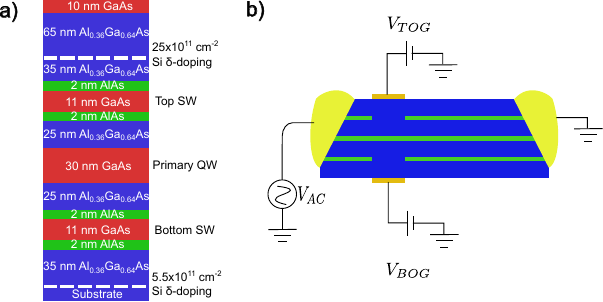}
    \caption{\raggedright \textbf{(a)} AlGaAs/GaAs screening well heterostructure layer sequence. Two 11~nm thick GaAs screening wells flank the primary 30~nm GaAs quantum well. \textbf{(b)} Schematic of the gates proximal to the ohmic contacts used to limit transport to the primary quantum well. The three quantum wells are in contact with an ohmic contact held at ground potential, but only the primary quantum well is connected to the excitation voltage. The top (bottom) screening well is isolated using a top (bottom) gate set at $V_{TOG}$ ($V_{BOG}$). This device design allows the screening wells to be held at a fixed potential while transport occurs only through the primary quantum well.
 }
    \label{FigS1}
\end{figure}
\par
The AlGaAs/GaAs heterostructure used in this study was grown using molecular beam epitaxy. The layer sequence is shown in Fig.~S1. The primary quantum well has a density of $n=1.0\times10^{11} \text{ cm}^{-2}$ and mobility of $\mu=9\times10^6 \text{ cm}^2/\text{Vs}$. The heterostructure incorporates additional screening wells separated by a $25~ \text{nm}$ from the primary quantum well. During device operation, transport occurs only through the primary quantum well. The screening wells are isolated from the ohmic contacts using top and bottom metallic gates. These gates are located near the ohmic contacts far from the mesoscopic region that defines the Fabry-P{\'e}rot interferometer. Negative voltage is applied to deplete the screening wells only in a narrow region around the ohmic contacts as shown in Fig.~S1b.
\par

\begin{figure}[htp!]
    \centering
    \includegraphics[width = 0.8\linewidth]{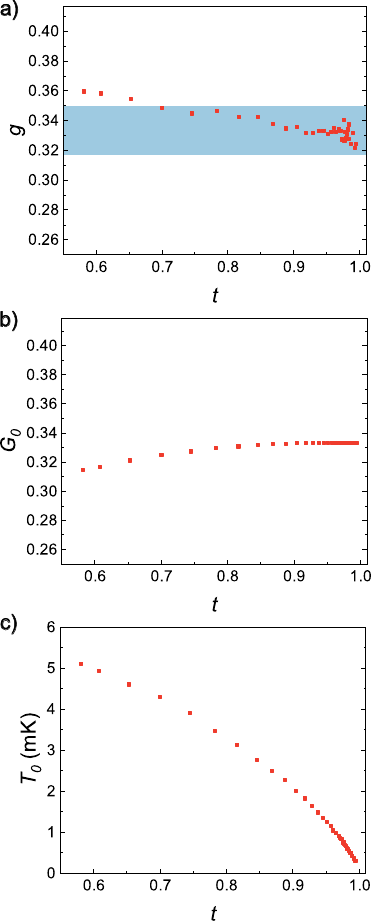}
    \caption{\raggedright \textbf{(a)} $g$ extracted from fitting the differential conductance data in Fig.~2b, plotted as a function of transmission. The translucent blue box indicates a 5\% deviation from $g = 1/3$. \textbf{(b)} $G_0$ extracted from fitting the differential conductance data in Fig.~2b, plotted as a function of transmission. The extracted $G_0$ remains extremely close to $G_0 = 1/3$ at high transmission, only showing deviation for $t\leq0.80$. \textbf{(c)} $T_0$ extracted from fitting the differential conductance data in Fig.~2b plotted as a function of transmission. $T_0$ is a measure of the tunneling strength between the edge modes and monotonically increases with decreasing $t$, as expected.}
    \label{FigS2}
\end{figure}

\subsection*{Description of data analysis}

In the limit of weak backscattering, $G_t$ is given by equations (\ref{eq.1}) and (\ref{eq.1a}) where the beta, digamma and gamma functions in equation (\ref{eq.1a}) are defined by: 
\begin{align}\label{eqS1}
    &B(z_1, z_2) = \frac{\Gamma(z_1)\Gamma(z_2)}{\Gamma(z_1+z_2)}\\[5pt]
    &\psi(z) = \frac{d}{dz}\ln\Gamma(z)\\[5pt]
    &\Gamma(z) = \int_0^{\infty}{t^{z-1}e^{-t}\,dt}
\end{align}

\par

The differential conductance data sets displayed in Fig.~2b are fitted to the equation $G=G_0-G_t$. We perform fits to the differential conductance data with $e^*=e/3$ and $T_e=$~34mK. These parameters are fixed by independent measurements described in the main text. We fit the data using the \textit{scipy.optimize.curve\_fit} function in Python. This function minimizes the sum of squared residuals and has as its argument the raw data for each differential conductance data set with free parameters $g$, $G_0$, and $T_0$. The output of the function is the set of three optimal parameters and the covariance matrix. The square roots of the diagonal elements of the covariance matrix are the uncertainties of each parameter. These uncertainties are defined so that the reduced $\chi^2$ of the fit is equal to one. As an example, the raw data in Fig.~2c has 498 data points. We input an array with these points into the \textit{scipy.optimize.curve\_fit} function. For this particular case, the output of the fit in Fig.~2c is: $g=0.334\pm0.001$, $T_0=0.001339\pm0.000001$ and $G_0=0.33353\pm0.00002$ where $\pm$ is given by the corresponding square of the diagonal element in the covariance matrix corresponding to each parameter. 
 
\par
The output for the fits of the data sets displayed Fig.~2b of the main text are shown in Fig.~S2. Our extractions for $g$ and $G_0$ are stable in the weak backscattering limit and are consistent with the theoretical description of the CLL \cite{Wen1991}. A $\pm5\%$ band around $g=1/3$ is shown in Fig.~S2a. Deviations from $g=1/3$ are only realized for lower transmissions $t\leq0.7$. The extracted values for $T_0$ is a monotonically decreasing function of $t$ controlled by $V_{QPC}$, as expected for our QPC.
\color{black}
\subsection*{Coulomb Blockade Thermometry}
\begin{figure}[htp!]
    \centering
    \includegraphics[width = 0.8\linewidth]{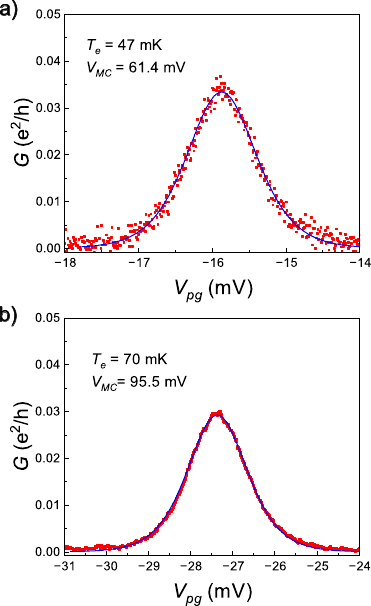}
    \caption{\raggedright \textbf{(a)} Conductance in the Coulomb blockade regime as a function of plunger gate voltage. For this measurement, an excitation voltage $V_{MC} = 61.4 \text{mV}$ is applied to the mixing chamber heater, and the system has been allowed to equilibrate. \textbf{(b)} Same as \textbf{(a)} but with $V_{MC} = 95.5\text{mV}$.
 }
    \label{FigS3}
\end{figure}
The QPC used for our measurements of tunneling conductance is part of a 1$\mu$m$^2$ Fabry-P{\'e}rot interferometer. We measure the electron temperature {\it in operando} by operating the interferometer in the Coulomb blockade regime. We measured the charging energy of this quantum dot to be $e^2/C \approx 91 \mu\text{eV}$ and the lever arm $\alpha \approx 0.016$. The zero bias conductance $G$ as a function of the plunger voltage $V_{pg}$ can then be fitted to $G \propto cosh^{-2}(\alpha V_{pg}/2k_bT)$ to extract the temperature under various measurement conditions, as shown in Fig.~S3. This technique allows for accurate temperature measurements of the electron temperature in an operating device.

\subsection*{Additional Data Sets}

\begin{figure}[htp!]
    \centering
    \includegraphics[width = 0.8\linewidth]{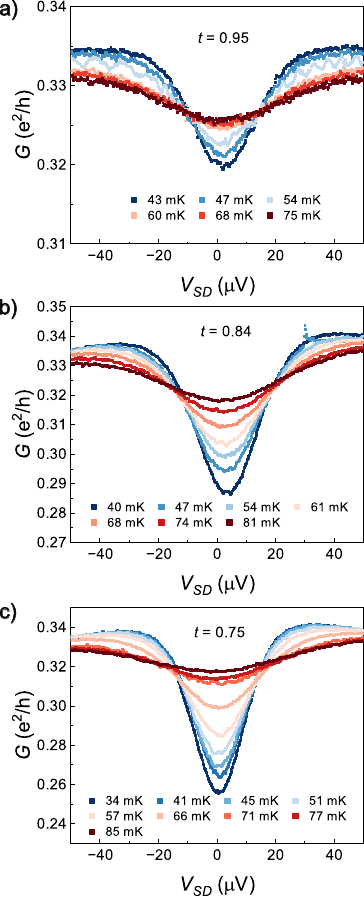}
    \caption{\raggedright \textbf{(a)} Temperature dependence of the differential conductance for QPC transmission $t=0.95$. Temperatures listed in the legend are mixing chamber temperatures at which each data set was collected. \textbf{(b)} Same as \textbf{(a)} but with $t=0.84$. \textbf{(c)} Same as \textbf{(a)} but with $t=0.75$.
 }
    \label{FigS4}
\end{figure}
\par
The data displayed in Fig.~2d of the main text shows the temperature dependence of the differential conductance at QPC transmission $t=0.9$. We measured the temperature dependence at several additional values of QPC transmission in the weak backscattering limit. This data is displayed in Fig.~S4.
\par
Fig.~S5 displays $R_D$ and $R_{xy}$ between $\nu=1$ and $\nu=1/3$ taken while sweeping the magnet up to $\nu=$1/3. The nonmonotonic evolution of $R_D$ and $R_{xy}$ is associated with the redistribution of charge between the three quantum wells when $B\approx7$~T. Both $R_D$ and $R_{xy}$ are well quantized at $\nu=1/3$.
\begin{figure}[htp!]
    \centering
    \includegraphics[width = 0.8\linewidth]{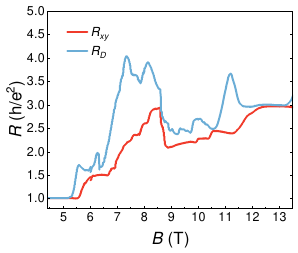}
  \caption{\raggedright Simultaneous measurement of bulk $R_{xy}$ and $R_D$ as a function of magnetic field from the center of $\nu = 1$ to just above $\nu = 1/3$. Well quantized plateau in both $R_D$ and $R_{xy}$ are observed around $B \approx 12.8\,\text{T}$
 }
    \label{FigS4}
\end{figure}
\end{appendices}

\end{document}